\documentclass[12pt,preprint]{emulateapj}
\def\apj{{ApJ}}
\def\apjl{{ApJL}}
\def\mnras{{ MNRAS}}
\def\aa{{ A\&A}}

\def\be{\begin{equation}}
\def\ee{\end{equation}}
\def\bea{\begin{eqnarray}}
\def\eea{\end{eqnarray}}
\usepackage{graphicx}
\usepackage{amsmath}
\usepackage[colorlinks,
            linkcolor=blue,
            anchorcolor=blue,
            citecolor=blue]{hyperref}

\usepackage{color,txfonts}
\begin{document}

\title{HESS J1640-465 -- a Gamma-ray emitting pulsar wind nebula ?}

\author{Yu-Liang Xin\altaffilmark{1,2}, Neng-Hui Liao\altaffilmark{1,3}, Xiao-Lei Guo\altaffilmark{1,3}, 
Qiang Yuan\altaffilmark{1,3}, Si-Ming Liu\altaffilmark{1,3}, Yi-Zhong Fan\altaffilmark{1,3}, Da-Ming Wei\altaffilmark{1,3}}
\email{E-mail: yuanq@pmo.ac.cn (QY); liusm@pmo.ac.cn (SML); dmwei@pmo.ac.cn (DMW)}
\altaffiltext{1}{Key laboratory of Dark Matter and Space Astronomy, Purple Mountain Observatory, Chinese Academy of Sciences, Nanjing 210008, China;}
\altaffiltext{2}{University of Chinese Academy of Sciences, Yuquan Road 19, Beijing, 100049, China;}
\altaffiltext{3}{School of Astronomy and Space Science, University of Science and Technology of China, Hefei, Anhui 230026, China}

\begin{abstract}

HESS J1640-465 is an extended TeV $\gamma$-ray source and 
its $\gamma$-ray emission whether from the shell of a supernova remnant (SNR)
or a pulsar wind nebula (PWN) is still under debate.
We reanalyze the GeV $\gamma$-ray data in the field of HESS J1640-465 using eight years of Pass 8 data
recorded by the Fermi Large Area Telescope. 
An extended GeV $\gamma$-ray source positionally coincident with HESS J1640-465 is found.
Its photon spectrum can be described by a power-law with an index of $1.42\pm0.19$ 
in the energy range of 10-500 GeV, and smoothly connects with the TeV spectrum of HESS J1640-465.
The broadband spectrum of HESS J1640-465 can be well fit by a leptonic model
with a broken power-law spectrum of electrons with an exponential cut-off at $\sim$ 300 TeV.
The spectral properties of HESS J1640-465 are broadly consistent with the characteristics of other sources identified as PWNe,
such as the correlations between high-energy luminosity ratios and the physical parameters of pulsar,
including spin-down luminosity $\dot{E}$ and characteristic age $\tau_c$.
All these pieces of evidence support that the $\gamma$-ray emission of HESS J1640-465 
may originate from the PWN powered by PSR J1640-4631 rather than the shell of the SNR G338.3-0.0.

\end{abstract}


\setlength{\parindent}{.25in}

\section{Introduction}

Since the beginning of operations in 2004, the High Energy Stereoscopic System (HESS) 
opened a new era for studies of the extremely energetic phenomena of the Universe,
yielding a large number of very high energy (VHE; $>$ 100 GeV) $\gamma$-ray 
sources. A large portion of the Galactic sources had been firmly identified as supernova remnants (SNRs) 
and pulsar wind nebulae (PWNe), with the latter being the most numerous class.
While the radio to the non-thermal X-ray emission
is typically produced by synchrotron radiation of high energy electrons, 
the $\gamma$-ray emission can either be due to the inverse Compton scattering (ICS)/bremsstrahlung process of 
electrons (leptonic process) or the decay of neutral pions produced by inelastic $pp$ collisions (hadronic process).

For dynamically older SNRs interacting with molecular clouds, 
their $\gamma$-ray emission is typically believed to be of hadronic origin, 
such as IC 443 \citep{Ackermann2013}, W44 \citep{AGILE2011,Ackermann2013},
Puppis A \citep{Hewitt2012, Xin2017} and so on.
While for dynamically younger SNRs with fast efficient shocks, such as 
RX J1713-3946 \citep{Yuan2011,Zeng2017,Abdalla2016a}, 
RCW 86 \citep{Yuan2014,Ajello2016},
and HESS J1731-347 \citep{Abramowski2011, Guo2018, Condon2017},
the $\gamma$-ray emission is usually thought to be from the leptonic process.
For PWNe, however, the $\gamma$-ray emission is mostly contributed by the leptonic process,
such as Vela X \citep{Abdo2010a, Aharonian2006a, Abramowski2012},
MSH 15-52 \citep{Abdo2010b, Aharonian2005}, 
and HESS J1825-137 \citep{Aharonian2006b, Grondin2011}.

HESS J1640-465 is one of the TeV sources discovered by the HESS survey of the inner 
Galaxy \citep{Aharonian2006c}. It was found to be marginally extended and spatially 
consistent with the broken shell-type SNR G338.3-0.0 \citep{Green2014}.
{\em XMM-Newton} observations revealed an extended, hard-spectrum X-ray emitting source at
the centroid of HESS J1640-465 \citep{Funk2007}.
This X-ray source was considered to be the counterpart of HESS J1640-465, which is expected to be a pulsar 
wind nebula (PWN) associated with G338.3-0.0. Follow-up observations with {\em Chandra}
identified a point source surrounded by diffuse X-ray emission \citep{Lemiere2009}.
The point source was suggested to be a pulsar, and the diffuse emission was its 
associated PWN. Meanwhile, \citet{Lemiere2009} constrained the distance of the system to be
between 8.5 and 13 kpc using HI absorption features.
The multifrenquency radio observations did not detect the radio pulsations towards the point
source \citep{Giacani2008, Castelletti2011}. Also no radio counterpart to the postulated X-ray PWN was 
found and only upper limits on the radio fluxes in this region were derived.
Later, \citet{Gotthelf2014} discovered X-ray pulsations from the point source using the 
{\em Nuclear Spectroscopic Telescope Array (NuSTAR)} observations and eventually identified it as a pulsar. 
The pulsar, PSR J1640-4631, has a spin period of 206 ms, a period derivative of
$\dot{P} = 9.758(44) \times 10^{-13}~{\rm s}~{\rm s}^{-1}$ 
and spin-down luminosity of $\dot{E} = 4.4 \times 10^{36}~{\rm erg}~{\rm s}^{-1}$.
The corresponding characteristic age of the pulsar is $\tau_c \equiv P/2\dot{P} = 3350 ~{\rm yr}$,
and the surface dipole magnetic field strength is 
$B_s = 3.2 \times 10^{19} (P\dot{P})^{1/2} ~{\rm G}$ = $1.4 \times 10^{13}~{\rm G}$.
Using the X-ray timing observations by {\em NuSTAR}, \citet{Archibald2016} measured its braking index to be 
$n = 3.15 \pm 0.03$.
The broadband spectrum fitting revealed that the TeV emission from HESS J1640-465 can be
explained by an evolving PWN powered by PSR J1640-4631 \citep{Gotthelf2014}.

The Fermi Large Area Telescope (Fermi-LAT) observations revealed a high energy $\gamma$-ray source,
1FGL J1640.8-4634, with a spectral index $\Gamma = 2.30 \pm 0.09$, 
which is positionally coincident with HESS J1640-465 \citep{Slane2010}.
The multi-wavelength emission from HESS J1640-465 can be interpreted by a PWN 
with a low magnetic field strength.
However, the follow-up observations with HESS showed that the TeV $\gamma$-ray emission of 
HESS J1640-465 significantly overlaps with the northwestern part of the shell of 
SNR G338.3-0.0 \citep{Abramowski2014a}. 
Furthermore, the TeV $\gamma$-ray spectrum can smoothly
connect with the relatively soft GeV $\gamma$-ray spectrum reported by \citet{Slane2010} with a high-energy cutoff. 
Considering the TeV morphology and the overall $\gamma$-ray spectrum,
\citet{Abramowski2014a} argued that a PWN scenario was hard to explain the $\gamma$-ray
spectrum of HESS J1640-465 while a hadronic scenario was favored.
For the hadronic scenario,
the protons can be accelerated in the shell of SNR G338.3-0.0 and the $\gamma$-ray emission
is produced by inelastic collisions between accelerated protons 
and dense gas associated with the nearby HII complex, G338.4+0.1.
\citet{Lemoine-Goumard2014} reanalyzed the region of HESS J1640-465 with more data of Fermi-LAT,
giving a slightly harder spectrum with an index of $1.99\pm0.04\pm0.07$ than that reported by \citet{Slane2010}. 
Although the flat $\gamma$-ray spectrum strengthens the hadronic scenario, 
the PWN scenario can not be ruled out \citep{Lemoine-Goumard2014,Gotthelf2014}.
Therefore, the nature of HESS J1640-465 is still not clear.

In the proximity of HESS J1640-465, another source, HESS J1641-463, 
was also detected with TeV $\gamma$-ray emission \citep{Abramowski2014b}.
HESS J1641-463 shows a hard TeV spectrum with an index of $\sim$ 2.0 and no obvious evidence of a high-energy cutoff.
HESS J1641-463 is positionally coincident with the radio SNR G338.5+0.1 \citep{Abramowski2014b}.
No conclusive counterparts in the X-ray band except several weak sources was detected by {\em XMM-Newton} and {\em Chandra}.
\citet{Lemoine-Goumard2014} reported the detection of the GeV $\gamma$-ray emission in the direction of HESS J1641-463 
using Fermi-LAT data. However, the GeV spectrum is relatively soft with an index of $\sim$ 2.47, 
which is quite different from the hard 
spectrum in the TeV band. This result indicates that the GeV and TeV emissions from HESS J1641-463 may have
different origins \citep{Tang2015,Lau2017}.

In this paper, we carry out a detailed analysis of GeV $\gamma$-ray emission of HESS J1640-465 with eight years of Fermi-LAT Pass 8 data.
In Section 2, we describe the data analysis routines and results. The nature of HESS J1640-465 and HESS J1641-463 based on  
the multi-wavelength observations is discussed in Section 3. Finally, we conclude our study in Section 4.

\section{Data analysis}

\subsection{Data reduction}
In the following analysis, we use the latest Fermi-LAT Pass 8 data collected from 
August 4, 2008 (Mission Elapsed Time 239557418) to August 4, 2016 (Mission Elapsed Time 491961604).
To avoid a too large point-spread function (PSF) in the low energy band, we only select events with
energies between 1 GeV and 500 GeV. The ``Source'' event class (evclass=128 \& evtype=3) is selected 
with the maximum zenith angle of $90^\circ$ to minimize the contamination from the Earth Limb.
The region of interest (ROI) is a square region of $10^\circ \times 10^\circ$ centered at
the position of 3FGL J1640.4-4634c (R.A.$=250.118^\circ$, Dec.$=-46.580^\circ$),
which was regarded as the GeV counterpart of HESS J1640-465 in the third Fermi-LAT source catalog \citep[3FGL;][]{Acero2015}.

The data analysis is performed using the standard LAT analysis software, 
{\it ScienceTools} version {\tt v10r0p5}\footnote {http://fermi.gsfc.nasa.gov/ssc/data/analysis/software/}, 
available from the Fermi Science Support Center. The binned likelihood analysis method and the instrument 
response function (IRF ) of ``P8R2{\_}SOURCE{\_}V6'' are adopted. For the background subtraction,
the Galactic diffuse emission ({\tt gll\_iem\_v06.fits}) and the isotropic background 
({\tt iso\_P8R2\_SOURCE\_V6\_v06.txt}\footnote {http://fermi.gsfc.nasa.gov/ssc/data/access/lat/BackgroundModels.html})
are included in the source model, as well as all sources in the 3FGL catalog within a radius of $18^\circ$ from the ROI center.

\subsection{Nature of 3FGL J1640.4-4634c}

\begin{table}[!htb]
\centering
\normalsize
\caption {Coordinates and TS values of the nine newly added point sources}
\begin{tabular}{cccc}
\hline \hline
Name & R.A. [deg] & Dec. [deg] & TS \\
\hline
NewPts1  & $245.889$ & $-49.5941$ & $167.118$ \\
NewPts2  & $247.357$ & $-48.2624$ & $135.538$ \\
NewPts3  & $248.777$ & $-48.1163$ & $127.615$ \\
NewPts4  & $248.737$ & $-47.4422$ & $126.384$ \\
NewPts5  & $249.535$ & $-47.3061$ & $110.339$ \\
NewPts6  & $252.838$ & $-44.2910$ & $78.069$ \\
NewPts7  & $253.859$ & $-46.4139$ & $69.584$ \\
NewPts8  & $251.508$ & $-45.5713$ & $45.905$ \\
NewPts9  & $253.083$ & $-45.3011$ & $41.398$ \\

\hline
\hline
\end{tabular}
\label{table:newpts}
\end{table}

\begin{table}[!htb]
\centering
\normalsize
\caption {Spectral parameters and TS values of sources A and B for different energy bands and point source assumptions}
\begin{tabular}{cccccc}
\hline \hline
1-10 GeV  & Spectral  & Photon Flux                        & TS         \\
          &  Index    & ($10^{-10}$ ph cm$^{-2}$ s$^{-1}$) & Value      \\
\hline
Source A & $2.61\pm0.14$  & $56.5\pm5.56$                     & 228.8     \\
Source B & 1.24 (fixed)   & $4.83\pm1.99$                     & 9.8       \\
\hline \hline
10-500 GeV & Spectral   & Photon Flux                        & TS         \\
           &   Index    & ($10^{-10}$ ph cm$^{-2}$ s$^{-1}$) & Value      \\
\hline
Source A & 2.61 (fixed)   & $1.49\pm0.39$                     & 22.8       \\
Source B & $1.24\pm0.22$  & $3.03\pm0.45$                     & 148.6     \\
\hline \hline
\end{tabular}
\label{table:SourceAB}
\end{table}

\begin{figure}[!htb]
\centering
\includegraphics[angle=0,scale=0.35,width=0.5\textwidth,height=0.38\textheight]{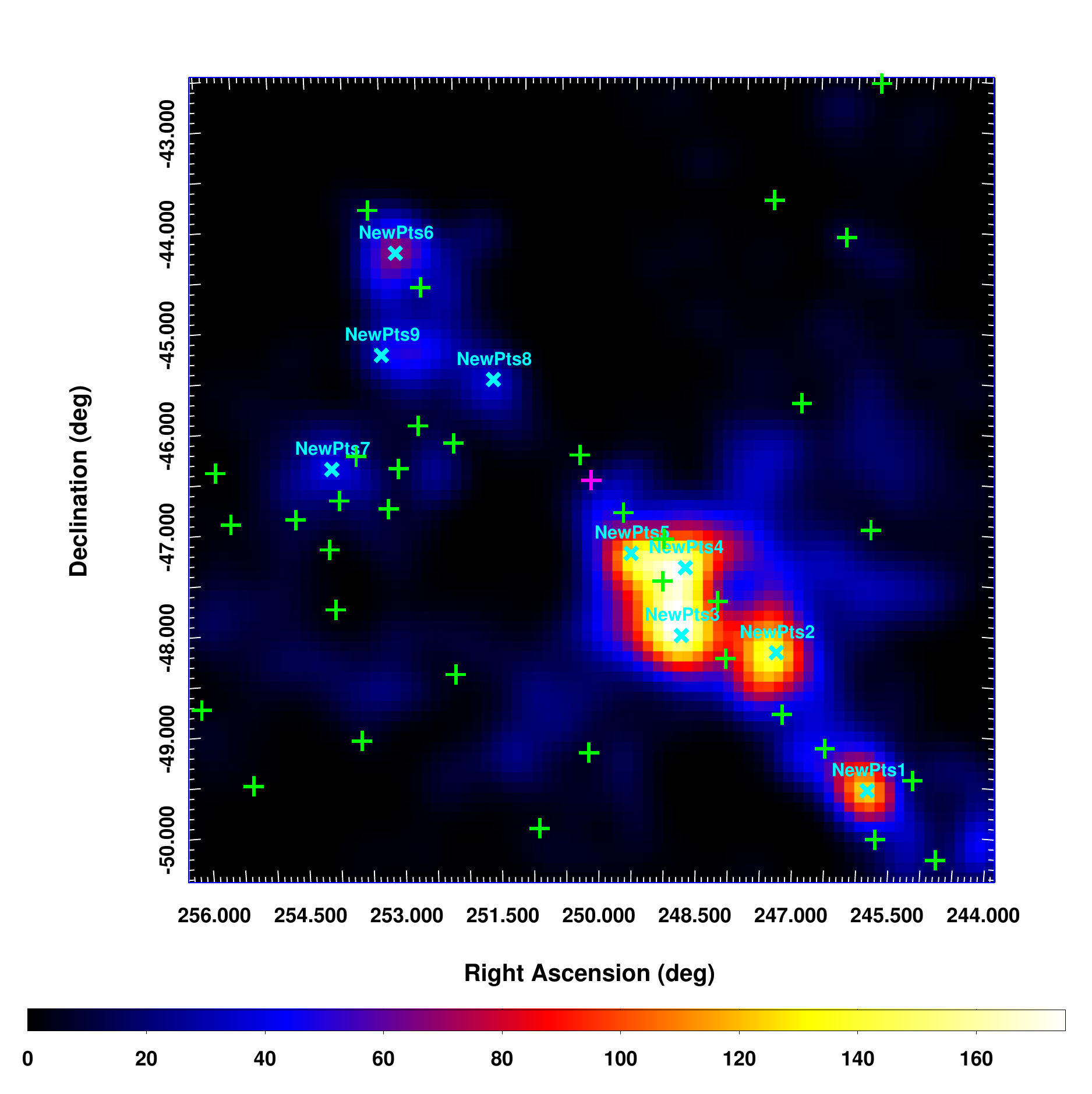}%
\\
\includegraphics[angle=0,scale=0.35,width=0.5\textwidth,height=0.3\textheight]{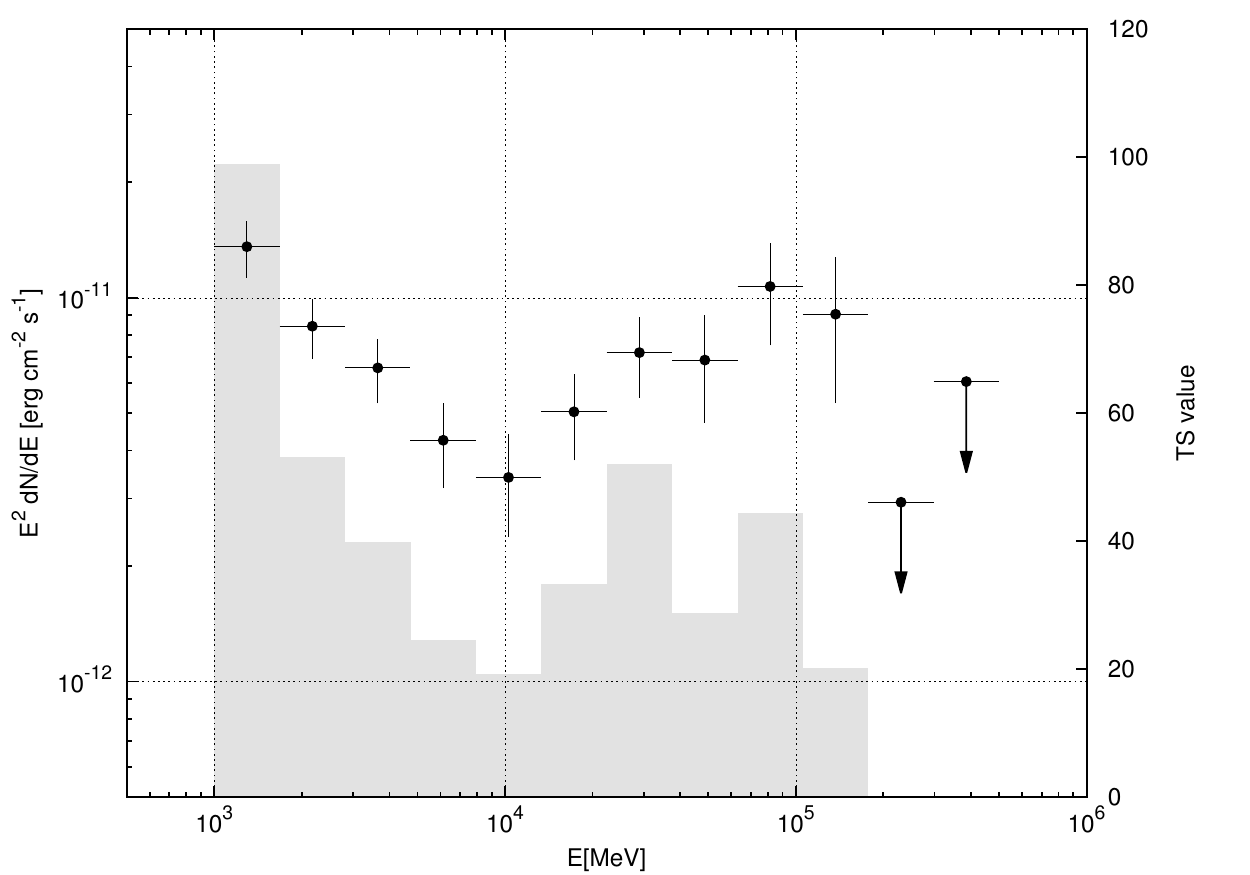}%
\hfill
\caption{Top: TS map for a $8.0^{\circ}\times8.0^{\circ}$ region centered at 3FGL J1640.4-4634c,
which is smoothed with a Gaussian kernel of $\sigma$ = $0.06^\circ$.
The TS map is for the model with only the 3FGL sources and the diffuse backgrounds.
The positions of the 3FGL sources \citep{Acero2015} are shown as the green pluses and 
the position of 3FGL J1640.4-4634c is marked by the magenta one.
The white crosses represent the best-fitting positions of the nine newly added point sources which are not included in 3FGL.
Bottom: SED of 3FGL J1640.4-4634c. The arrows indicate the 95\% upper limits
and the gray histogram denotes the TS value for each energy bin.}
\label{fig:tsmap1-sed-1-500}
\end{figure}

\begin{figure*}[!htb]
\centering
\includegraphics[angle=0,scale=0.35,width=0.5\textwidth,height=0.4\textheight]{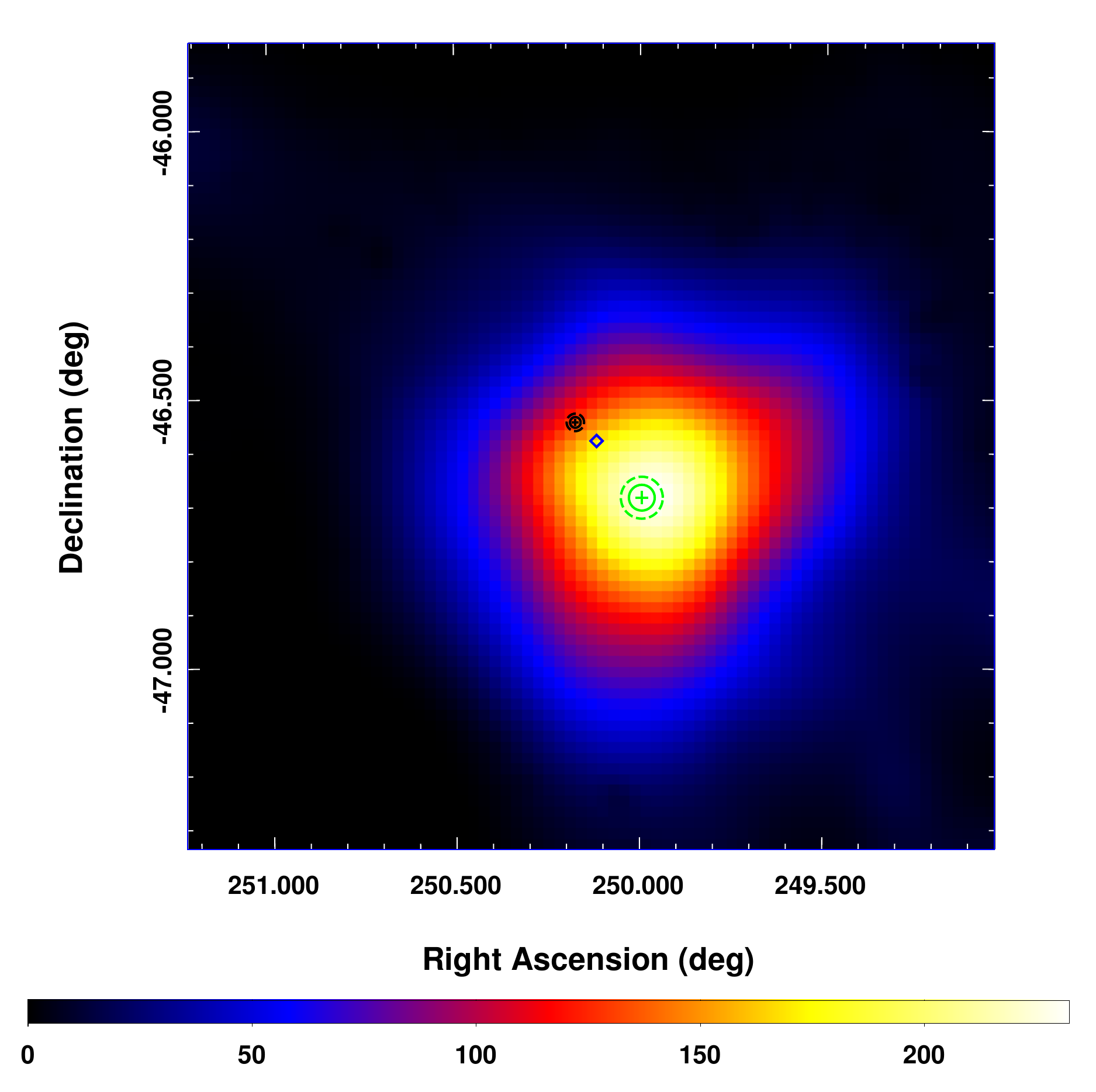}%
\includegraphics[angle=0,scale=0.35,width=0.5\textwidth,height=0.4\textheight]{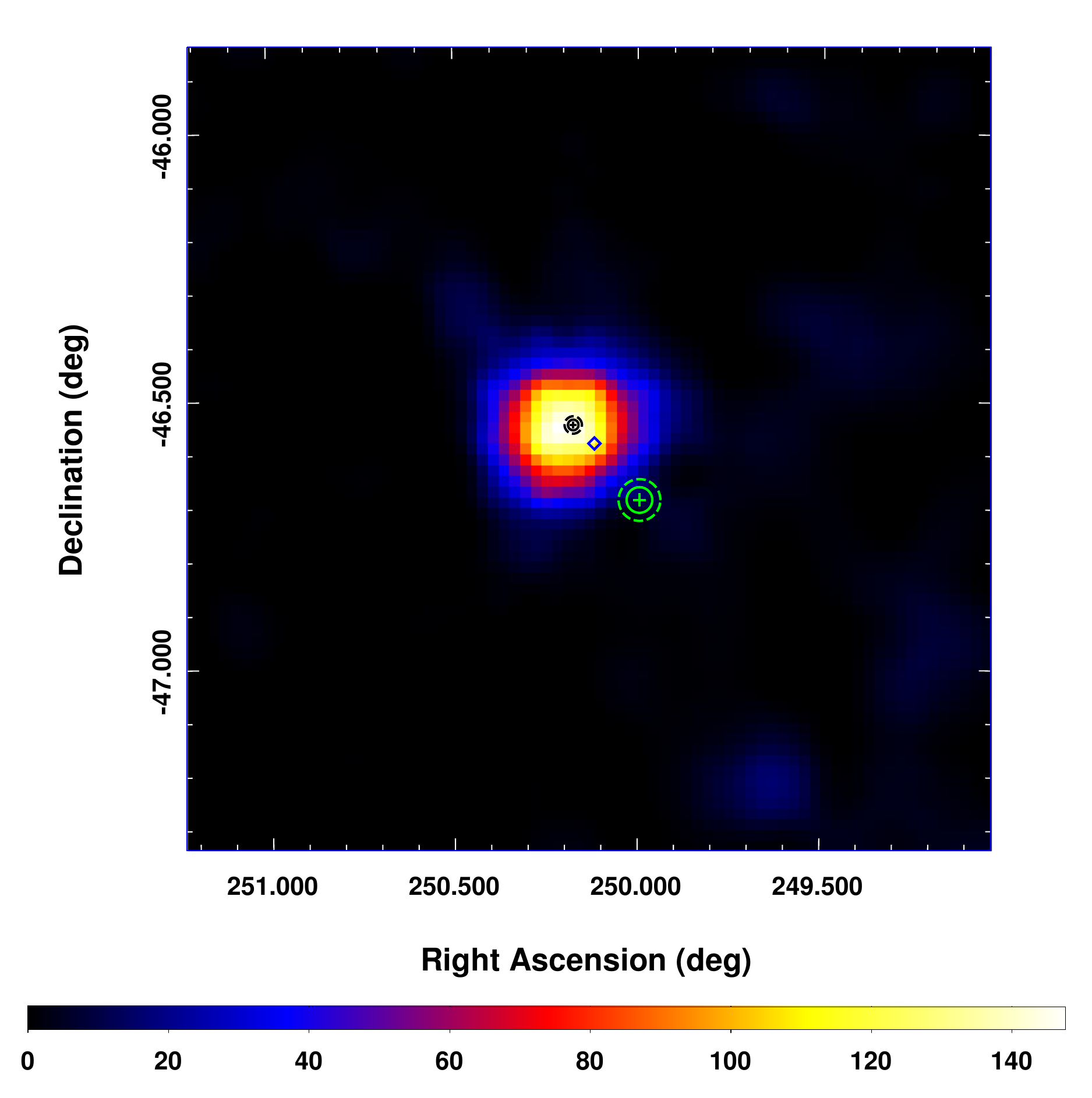}%
\hfill
\caption{TS maps of $1.5^{\circ}\times1.5^{\circ}$ region centered at 3FGL J1640.4-4634c.
The left panel is for photons from 1 GeV to 10 GeV and the right panel is for photons in the energy range of 10-500 GeV.
In the each panel, the position of 3FGL J1640.4-4634c provided by the 3FGL catalog \citep{Acero2015} is marked with the blue diamond.
The green and black pluses are the best-fitting position of Source A and B, respectively. 
And the corresponding 1$\sigma$ (solid) and 2$\sigma$ (dash) error circles of the fitting position 
are shown as the green and black circles.
The maps are smoothed with a Gaussian kernel of $\sigma$ = $0.04^\circ$.}
\label{fig:tsmap2}
\end{figure*}

During the likelihood analysis, the normalizations and spectral parameters of all sources within $5^\circ$ 
of 3FGL J1640.4-4634c, as well as the normalizations of the two diffuse backgrounds, are left free.
First, we create a Test Statistic (TS) map 
by subtracting the emission from the sources and backgrounds in the best-fit model 
with {\tt gttsmap}, which is shown in the top panel of Fig \ref{fig:tsmap1-sed-1-500}.
Some residual emission are shown in this TS map. 
Then we add additional point sources with power-law spectra in the model. 
The accurate positions of these sources obtained using the {\tt gtfindsrc} tool, 
together with their TS values, are listed in Table \ref{table:newpts}.
Next, we adopt the position of 3FGL J1640.4-4634c provided by the 3FGL catalog \citep{Acero2015} provisionally 
and investigate the spectrum of 3FGL J1640.4-4634c.
We bin the data into twelve equal logarithmic energy bins from 1 GeV to 500 GeV, 
and repeat the same likelihood fitting for each energy bin.
In the model, the normalization parameters of sources within $5^\circ$ around 3FGL J1640.4-4634c 
and the two diffuse backgrounds are left free, while all spectral indices except 3FGL J1640.4-4634c are fixed. 
The 95\% upper limits are calculated for energy bins with TS values smaller than 4.
The resulting spectral energy distribution (SED) is shown in the bottom panel of Fig \ref{fig:tsmap1-sed-1-500}.
And an obvious spectral upturn is shown in the SED of 3FGL J1640.4-4634c at an energy 
of about 10 GeV. 
To test whether the upturn spectrum is intrinsic or due to two overlapping sources,
we do the same likelihood fitting using the events with energies of 1-10 GeV and 10-500 GeV, respectively.
For each analysis, we create a TS map with all sources (except 3FGL J1640.4-4634c) included in the model, 
which are shown in the Fig \ref{fig:tsmap2}.
The TS maps show clear difference between two energy bands, 
and the centroids of emission in both energy bands deviate from that of 3FGL J1640.4-4634c. 
We thus expect that 3FGL J1640.4-4634c should consist of two different sources 
(labelled as ``Source A'' for the 1-10 GeV source and ``Source B'' for the 10-500 GeV source), 
and the source in the 3FGL catalog is simply the sum of these two sources.

We delete 3FGL J1640.4-4634c and add Source A and B as point sources in the model file. 
We assume that their spectra are power-laws, and then fit to the data again.
The fitting results in the 1-10GeV energy band show that the emission is dominated by Source A.
The TS value of Source B is found to be smaller than 10.
The spectrum of Source A is soft with an index of 2.6.
The best-fitting coordinate of Source A is 
R.A.$=249.995^\circ$, Dec.$=-46.6857^\circ$, with $1\sigma$ uncertainty of $0.024^\circ$. 
Contrary to that in the 1-10 GeV band, the emission from Source A in the 10-500 GeV band is relatively
weak with a TS value smaller than 25, while Source B has significant GeV $\gamma$-ray emission in the high energy band.
The spectral index of Source B is fit to be about 1.24, and the best-fit coordinate is
R.A.$=250.175^\circ$, Dec.$=-46.5457^\circ$, with $1\sigma$ uncertainty of $0.01^\circ$.
We then have an iteration of the above analysis with the best-fit coordinates of Source A/B, 
and fix the spectral index of Source B (Source A) in the 1-10 (10-500) GeV analysis.
The fitting results are shown in Table \ref{table:SourceAB}.
The angular separation between Source A and B is about $0.19^\circ$, 
which is slightly larger than the size of the PSF (68\% containment radius) 
of Fermi-LAT for photon energy above 10 GeV \citep{Atwood2013}.
It should be noted that Source B is spatially consistent with the point source 3FHL J1640.6-4633 in the third catalog of 
hard Fermi-LAT sources \citep[3FHL;][]{Ajello2017}, which was assumed to be the likely counterpart of SNR G338.3-0.0.

\begin{figure}[!htb]
\centering
\includegraphics[angle=0,scale=0.4,width=0.5\textwidth,height=0.4\textheight]{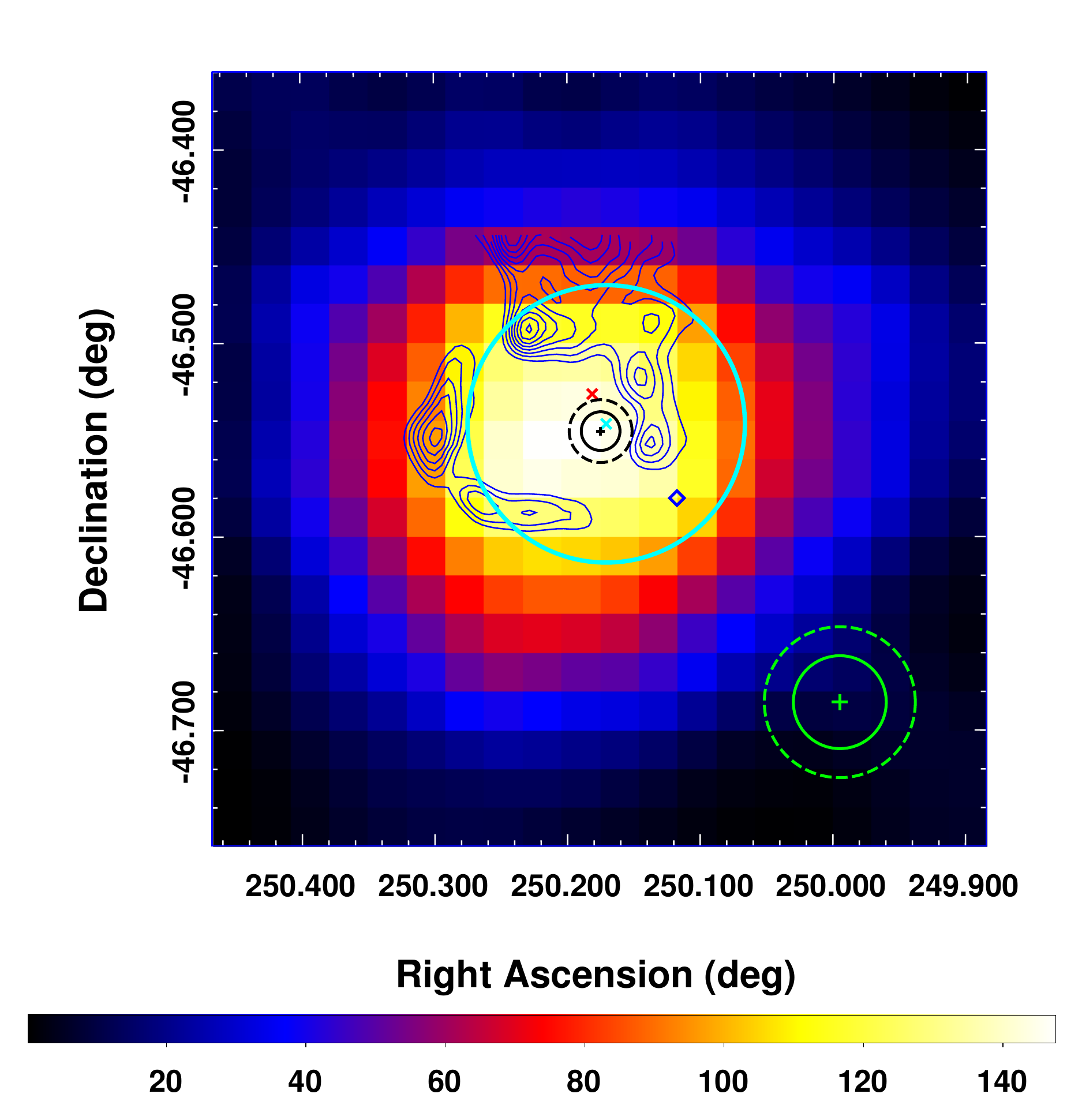}%
\hfill
\caption{Zoom-in of the TS map, for a region of $0.5^{\circ}\times0.5^{\circ}$ centered at the best-fitting position of Sourcc B.
The image was created using a grid of $0.02^{\circ}$ and smoothing is applied with a Gaussian kernel of $\sigma$ = $0.04^\circ$.
The best-fitting position of Source B is given by the black plus sign, together with the 1$\sigma$ and 2$\sigma$ error circles
indicating by the black solid and dashed circles, respectively. While for Source A, the best-fitting position and error circles are
marked in green. Blue contours represent the radio image of SNR G338.3-0.0 at 843 MHz from SUMSS \citep{Mauch2003}.
The cyan cross donates the best-fitting position of HESS J1640-465 in the TeV band 
and the intrinsic Gaussian width of $4.3'$ is shown as the cyan circle \citep{Abramowski2014a}.
The position of PSR J1640-4631 \citep{Gotthelf2014} is marked as the red cross.}

\label{fig:zoomin-tsmap-10-500GeV}
\end{figure}

To better understand the spatial correlation between Source A/B and the candidate counterparts in other wavelengths, 
especially HESS J1640-465, we create a zoom-in of the TS map for a region of $0.5^\circ$ $\times$ $0.5^\circ$ centered on Source B,
which is shown in Fig \ref{fig:zoomin-tsmap-10-500GeV}. 
Both HESS J1640-465 and SNR G338.3-0.0 are far beyond the 95\% error circle region of Source A.
Therefore, we suggest that Source A is neither associated with HESS J1640-465 nor SNR G338.3-0.0,
and we will take it as a background point source in the following analysis.
Source B is well coincident with HESS J1640-465, as well as the PWN powered by PSR J1640-4631 \citep{Gotthelf2014}, 
as shown in Fig \ref{fig:zoomin-tsmap-10-500GeV}.

\subsection{Spatial Analysis of Source B}

\begin{table*}[!htb]
\centering
\normalsize
\caption {spatial distribution analysis for Source B between 10 GeV and 500 GeV}
\begin{tabular}{cccccc}
\hline \hline
Spatial   & Radius ($\sigma$)  & Spectral & Photon Flux                         &  TS     & Degrees of \\
Template  &                    & Index    & ($10^{-10}$ ph cm$^{-2}$ s$^{-1}$)  & Value   & Freedom\\
\hline
Point Source & $--$          & $1.24\pm0.22$   & $3.03\pm0.45$    & 148.6   & 4 \\
Uniform disk & $0.05^\circ$  & $1.34\pm0.20$   & $4.97\pm0.71$    & 169.5   & 5 \\
             & $0.06^\circ$  & $1.35\pm0.20$   & $5.10\pm0.71$    & 170.7   & 5 \\
             & $4.3'^{\,\,\,\text{a}}$        & $1.37\pm0.19$   & $5.30\pm0.72$    & 172.4   & 5 \\
             & $0.08^\circ$  & $1.38\pm0.19$   & $5.39\pm0.72$    & 172.8   & 5 \\
             & $0.10^\circ$  & $1.40\pm0.19$   & $5.60\pm0.73$    & 173.2   & 5 \\
             & $0.12^\circ$  & $1.41\pm0.19$   & $5.74\pm0.73$    & 172.3   & 5 \\

2-D Gaussian & $0.05^\circ$  & $1.39\pm0.19$   & $5.52\pm0.73$    & 174.9   & 5 \\
             & $0.06^\circ$  & $1.40\pm0.19$   & $5.64\pm0.74$    & 174.8   & 5 \\
             & $4.3'^{\,\,\,\text{a}}$        & $1.42\pm0.19$   & $5.72\pm0.73$    & 174.3   & 5 \\
             & $0.08^\circ$  & $1.42\pm0.18$   & $5.78\pm0.74$    & 173.4   & 5 \\
             & $0.10^\circ$  & $1.43\pm0.18$   & $5.83\pm0.74$    & 171.9   & 5 \\
             & $0.12^\circ$  & $1.44\pm0.18$   & $5.87\pm0.74$    & 170.7   & 5 \\

\hline
\hline
\end{tabular}

\tablecomments{
\\
a) $4.3'$ is the best-fitting radius for TeV image of HESS J1640-465 \citep{Abramowski2014a}.
}
\label{table:spatial}
\end{table*}

As analysed by \citet{Abramowski2014a}, a symmetric two-dimensional (2-D) Gaussian profile 
with a width of $\sigma$ = (4.3 $\pm$ 0.2) arcmin can well describe the extended TeV $\gamma$-ray 
emission of HESS J1640-465. 
For the GeV $\gamma$-ray emission, \citet{Lemoine-Goumard2014} suggested that a Gaussian template could 
slightly improve the fit compared with a point-like source.
In \citet{Ackermann2017}, the GeV $\gamma$-ray emission was identified to be extended with a radius of $0.08^\circ \pm 0.02^\circ$ 
assuming a uniform disk. The TS value of the extension was found to be about 18 when using the {\em pointlike} analysis tool.

Here, we treat Source A as a point source and use the uniform disks centered at the best-fitting 
position with different radii, as well as the 2-D Gaussian profiles with different $\sigma$ as 
the spatial templates for Source B and re-do the fittings. The spectral index, 
photon flux and TS value for each spatial template are listed in Tabel \ref{table:spatial}.
The comparison between the different spatial models based on the Akaike information criterion \citep[AIC;][]{Akaike1974}\footnote{AIC is way of comparing the quality of a set of statistical models to each other to
seek a model that has a good fit to the data bu few parameters.} shows $\Delta$AIC = AIC$_{\rm point}$ - AIC$_{\rm ext}$ $\approx$ 24. Here AIC$_{\rm point}$ and AIC$_{\rm ext}$ are the AIC values for point-source and extended spatial templates, respectively.
This result shows the significant improvement when using an extended template instead of a point-source hypothesis.
This result is consistent with the TeV $\gamma$-ray extension of HESS J1640-465 \citep{Abramowski2014a}, 
which means that Source B is very likely the GeV counterpart of HESS J1640-465.
Considering the very close TS values for the uniform disks with different radii 
and the 2-D Gaussian profiles with different $\sigma$, we adopted the 2-D Gaussian profile with $\sigma$ = $4.3'$ 
(TeV spatial shape of HESS J1640-465) as the spatial template for Source B in the following spectral analysis.

\subsection{Spectral Analysis of Source B}

\begin{figure}[!htb]
\centering
\includegraphics[angle=0,scale=0.35,width=0.5\textwidth,height=0.3\textheight]{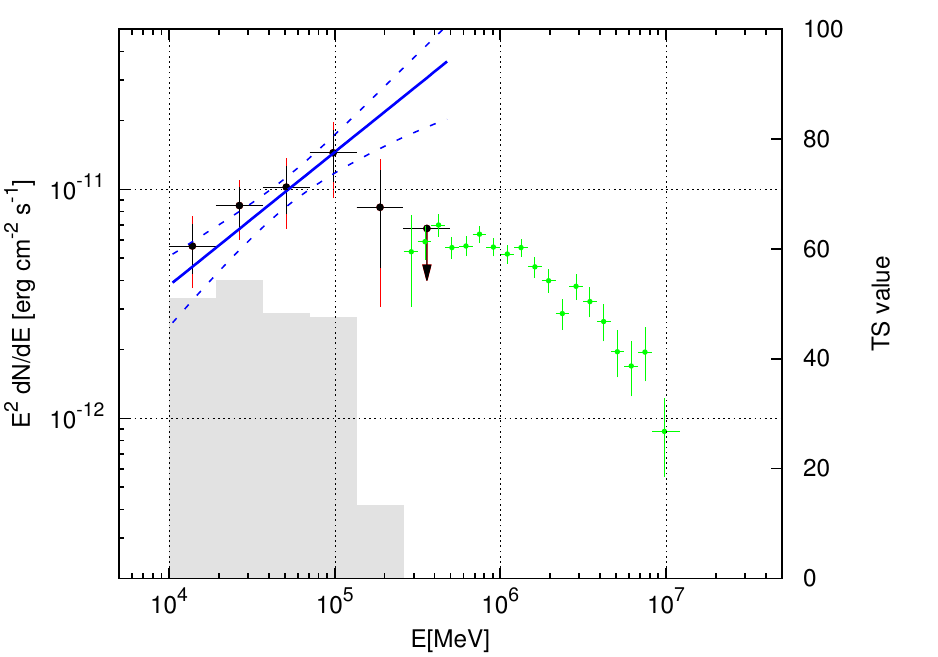}%
\hfill
\caption{SED of Source B. The black dots depict the results of Fermi-LAT data in the energy range of 10-500 GeV, 
with arrows indicating the 95\% upper limits.
The statistical errors are shown in black, while the red lines take both the statistical 
and systematic errors into consideration as discussed in Sect. 2.4.
The blue solid line is the best-fitting power-law spectrum in the energy range of 10-500 GeV, 
and the blue dashed line shows the 1$\sigma$ statistic error of the global fitting.
The gray histogram shows the TS value for each energy bin and the green dots in the TeV band are 
from the HESS observation \citep{Abramowski2014a}.} 
\label{fig:sed-10-500}
\end{figure}

With the spatial template of 2-D Gaussian profile with $\sigma$ = $4.3'$, 
the spectrum of Source B can be well fit by a power-law with an index of $1.42\pm0.19$. 
We also test a broken power-law spectrum and a power-law spectrum with an exponential cutoff for Source B,
but the results do not improve significantly.
The integral photon flux in the energy range from 10 GeV to 500 GeV is $(5.72\pm0.73)\times10^{-10}$ photon cm$^{-2}$ s$^{-1}$ 
with statistical error only.
Adopting a distance of $d = 10\mathrm{kpc}$ \citep{Lemiere2009,Abramowski2014a}, 
the $\gamma$-ray luminosity in the energy range of 10-500 GeV is $8.57\times 10^{35}\,(d/10\ {\rm kpc})^2$ erg~s$^{-1}$, 
which is several times higher than that of Crab Nebula \citep{Abdo2010c,Acero2013}.

To derive the $\gamma$-ray SED of Source B, we divided all of the data into six equal logarithmic energy bins 
from 10 GeV to 500 GeV.
For each energy bin, the spectral normalizations of all sources within $5^\circ$ from Source B and 
the two diffuse backgrounds are left free, 
while the spectral indices are fixed. For Source B, both the spectral normalization and index are left free.
For the energy bin with TS value smaller than 4, an upper limit at 95\% confidence level is calculated.
For the SED, two main systematic errors have been taken into account in the analysis:  
the imperfect modeling of the Galactic diffuse emission 
and the uncertainties of the effective area of the LAT.
The systematic error caused by the former is calculated by changing the best-fit normalization of the Galactic
diffuse model artificially by $\pm$6\% \citep{Abdo2010e}.
And the second one caused by the uncertainties of the effective area is estimated by using modified IRFs whose effective areas bracket the nominal ones
\footnote{https://fermi.gsfc.nasa.gov/ssc/data/analysis/scitools/Aeff$\_$Systematics.html}.
The SED and fit results are shown in Fig \ref{fig:sed-10-500}.
The SED shows that the Fermi-LAT data of Source B can connect with the TeV $\gamma$-ray spectrum of 
HESS J1640-465 smoothly,
which further supports Source B as the GeV counterpart of HESS J1640-465.



\section{Discussion}

From the above analysis, Source B with the extended $\gamma$-ray emission is positionally consistent with HESS J1640-465,
and the SED of Source B also connects with the TeV spectra of HESS J1640-465, which all support that Source B 
is the GeV $\gamma$-ray counterpart of HESS J1640-465 

However, considering the extended spatial morphology of the GeV emission, the coincidence between Source B and 
the PWN powered by PSR J1640-4631 or the shell of SNR G338.3-0.0 still can not be distinguished completely.
Nevertheless, the spatial coordinates between the pulsar PSR J1640-4631 and the $\gamma$-ray centroid of Source B
show the offset between them and such offset is very typical for $\gamma$-ray PWNe \citep{Abdalla2017}, 
such as HESS J1303-631 \citep{Abramowski2012}, 
HESS J1857+026 \citep{Rousseau2012} and HESS J1825-137 \citep{Grondin2011}.
The offset can be explained by either a PWN expansion into an inhomogenous medium 
or an asymmetric reverse shock interaction \citep{Blondin2001}.

We compare HESS J1640-465 with other SNRs 
which have the similar $\gamma$-ray spectra with HESS J1640-465\citep{Funk2015, Guo2017a},
such as RX J1713-3946 \citep{Abdo2011, Yuan2011, Zeng2017}, RX J0852-4622 \citep{Tanaka2011}, 
RCW 86 \citep{Yuan2014}, SN1006 \citep{Acero2010, Araya2012, Xing2016, Condon2017}, 
and HESS J1731-347 \citep{Abramowski2011, Guo2018, Condon2017}.
These SNRs are dynamically younger with fast efficient shocks.
And they have harder GeV spectra and are usually brighter in the TeV band than in the GeV band.
The $\gamma$-ray emission from these SNRs are usually believed to be leptonic \citep{Yuan2012, Funk2015, Yang2015}, 
although there are debates for some of them \citep{Inoue2012, Gabici2014}.
However, all these SNRs emit significant non-thermal X-ray emissions 
that are very different from HESS J1640-465 which no any non-thermal X-ray emission detected from the shell of SNR G338.3-0.0.

If the $\gamma$-ray emission of HESS J1640-465 comes from the hadronic process as considered in \citet{Abramowski2014a}, 
the spectral index of protons, 
which is approximately equal to that of the $\gamma$-ray emission\citep{Stecker1971}, should be harder than 1.5.
Such a spectrum of protons is difficult to be produced for traditional models of diffusive shock acceleration (DSA),
although several theoretical works under some very specific conditions in the ambient could give such hard $\gamma$-ray 
spectra \citep{Inoue2012,Gabici2014}.
Therefore, HESS J1640-465 is more likely to be the PWN powered by PSR J1640-4631.
Here, a simple leptonic model based on the multi-wavelength observations of HESS J1640-465 is discussed.

The radio counterpart of the PWN has not been detected and only an upper limit of its radio flux was derived 
by Giant Metrewave Radio Telescope (GMRT) observations \citep{Giacani2008, Castelletti2011, Gotthelf2014}.
Both {\em  Chandra} and {\em  NuSTAR} detected extended X-ray emission from the PWN.
Its X-ray spectra can be well fit by a power-law with an index of $2.2_{-0.4}^{+0.7}$ 
for the joint fitting to {\em  Chandra} and {\em  NuSTAR} data \citep{Lemiere2009, Gotthelf2014}.

In the modeling, a broken power-law spectrum with an exponential cutoff for electrons is assumed \citep{Bucciantini2011,Torres2014}, 
which is in the form of
\begin{eqnarray}
\dfrac{dN_e}{dE} \propto \dfrac{(E/E_{e,\rm br})^{-\gamma_{1}}}{1+(E/E_{e,\rm br})^{\gamma_{2}-\gamma_{1}}} \exp \left(- \dfrac{E}{E_{e, \rm cut}} \right)
\label{eq:e_spectra}
\end{eqnarray}
where $\gamma_{1}$ and $\gamma_{2}$ are the low-energy and high-energy spectral indices, respectively.
$E_{e, \rm br}$ and $E_{e, \rm cut}$ are the break and cut-off energies.
The low-energy spectral index, $\gamma_{1}$, is usually determined by the radio data. 
And for PWNe, the radio spectrum is flat with the radio index $\alpha$ ($\gamma_{1} = 2\alpha+1$) of 0-0.3, typically.%

As mentioned previously, The distance of HESS J1640-465 is adopted to be $d = 10 \mathrm{kpc}$ \citep{Lemiere2009,Abramowski2014a}.
The radius of HESS J1640-465 is about $r\approx12.5$ pc for an angular size of $4.3'$ at such a distance.
For the ICS process, three components of the interstellar radiation field are taken into account: 
the cosmic microwave background (CMB), 
the infrared blackbody component ($T_1=15$ K, $u_1=4~u_{\rm CMB}$ eV cm$^{-3}$) and 
the optical blackbody component ($T_2=5000$ K, $u_2=1.15~u_{\rm CMB}$ eV cm$^{-3}$) \citep{Slane2010,Gotthelf2014}.
The contribution from the bremsstrahlung is also considered 
assuming a gas density of $n_{\rm gas} = 1.0 ~{\rm cm}^{-3}$.
 
\begin{center}
\begin{figure*}[!htb]
\centering
\includegraphics[angle=0,scale=0.35,width=0.85\textwidth,height=0.5\textheight]{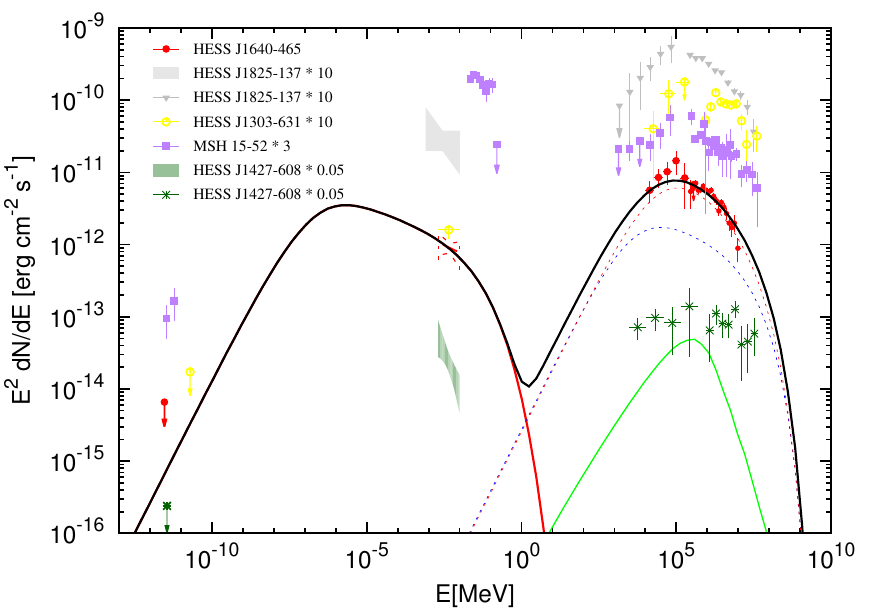}%
\hfill
\caption{Leptonic model for HESS J1640-465 as a PWN. 
The red arrow of HESS J1640-465 is the radio upper limit calculated from \citet{Gotthelf2014},
and the X-ray data marked by the red butterfly is taken from {\em  Chandra} 
and {\em  NuSTAR} observations \citep{Lemiere2009, Gotthelf2014}.
The radio and X-ray emissions are dominated by the synchrotron component, shown as the red solid line.
The $\gamma$-ray emission has contributions from the ICS (dashed lines) and bremsstrahlung (green solid lines) components.
The ICS emission includes three components from CMB (blue dashed), infrared (red dashed), 
and optical (too low to shown in this figure) radiation fields. The sum of the different radiation components for HESS J1640-465
are shown as the black solid line. The multi-wavelength data of HESS J1825-137 \citep{Gaensler2003, Uchiyama2009, Aharonian2006b, Grondin2011}, 
HESS J1303-631 \citep{Abramowski2012, Acero2013}, MSH 15-52 \citep{Gaensler1999, Gaensler2002, Forot2006, Aharonian2005, Abdo2010b},
and HESS J1427-608 \citep{Murphy2007, Fujinaga2013, Guo2017a} are also shown for comparison. 
And the energy fluxes of them are scaled upward for the sake of clarity as shown in the legend.}
\label{fig:leptonic}
\end{figure*}
\end{center}

The broadband spectrum and best fit model are shown in Fig. \ref{fig:leptonic}.
For the leptonic model, the spectral indices of the electrons, $\gamma_{1}$ and $\gamma_{2}$, are fit to be about
1.3 and 3.3, and the break and cutoff energies of electrons are about 1 TeV and 300 TeV, respectively.
The magnetic field strength of $\sim 5.5 \mu G$ and the total energy of electrons above 1 GeV of $3.2\times10^{48}$~erg 
are needed to explain the flux in the radio and X-ray bands.
Such a value of the magnetic field strength is also typical for PWNe \citep{Acero2013,Torres2014}.
The synchrotron radiation loss timescale is calculated by \citep{Parizot2006,Funk2015}
\begin{eqnarray}
t_{\rm syn} = 1.2 \times 10^{8} \left(\dfrac{B}{10 \mu \rm G}\right)^{-2} \left(\dfrac{E_{e, \rm br}}{\rm GeV}\right)^{-1} {\rm yr}
\label{eq:syn-loss-timescale}
\end{eqnarray}
Based on the values of the break energy and the magnetic field strength in the model, 
the synchrotron radiation loss timecale is much larger than the characteristic age of PSR J1640-4631 ($\sim$ 3000 yrs),
which indicates that the spectral break 
may be intrinsic for the electron spectrum injected in the PWN\citep{deJager2008}
instead of causing by radiative loss.
In addition, as suggested by \citet{Abdo2010b}, the spectral break of electrons 
may correspond to the energy scale for the switch of acceleration mechanisms, 
e.g. from the magnetic reconnection \citep{Zenitani2001} 
to the first-order Fermi acceleration \citep{Blandford1987}.

\citet{Gao2017} introduced a mean rotation energy conversion coefficient 
and combined the equation of state with the high-energy and timing observations of PSR J1640-4631
to estimate the initial spin period and the momentum of inertia of the pulsar.
They gave an initial spin period of $P_0\sim(17-44)~{\rm ms}$, 
which corresponds to a momentum of inertia of $I\sim(0.8-2.1)\times10^{45} ~{\rm g}~{\rm cm}^2$ \citep{Gao2017}.
Adopting such values of $P_0$ and $I$, the total energy of PSR J1640-4631 
is estimated to be $(2.1-5.5)\times10^{49}$~erg.
About 6\%-15\% of the pulsar's spin-down energy is required to be 
converted to accelerated electrons to explain the $\gamma$-ray emission, 
which is in agreement with the results given by \citet{Slane2010}.
And such efficiency is also plausible for PWNe\citep{Mattana2009,Acero2013}.


Also shown in Fig. \ref{fig:leptonic} are multi-wavelength data of several PWNe for comparison. 
Their fluxes are re-scaled for a better view.
In the GeV-TeV $\gamma$-ray band, HESS J1640-465 has similar $\gamma$-ray spectra with 
HESS J1825-137 \citep{Aharonian2006b, Grondin2011}, HESS J1303-631 \citep{Abramowski2012, Acero2013} 
and MSH 15-52 \citep{Aharonian2005, Abdo2010b}, all of which exhibit a hard GeV spectrum and peak at $\sim$ 100 GeV.
Meanwhile, the X-ray spectra of the three sources \citep{Gaensler2003, Uchiyama2009, Aharonian2006b, Forot2006}
also show the same spectral behaviour as HESS J1640-465.
However, the X-ray luminosity of MSH 15-52 is about one order of magnitude higher than its $\gamma$-ray luminosity,
which is opposite with that of HESS J1825-137, HESS J1303-631 and HESS J1640-465. 
This can be explained by a higher magnetic field strength in MSH 15-52 \citep{Abdo2010b}.

\begin{figure*}[!htb]
\centering
\includegraphics[width=0.5\textwidth]{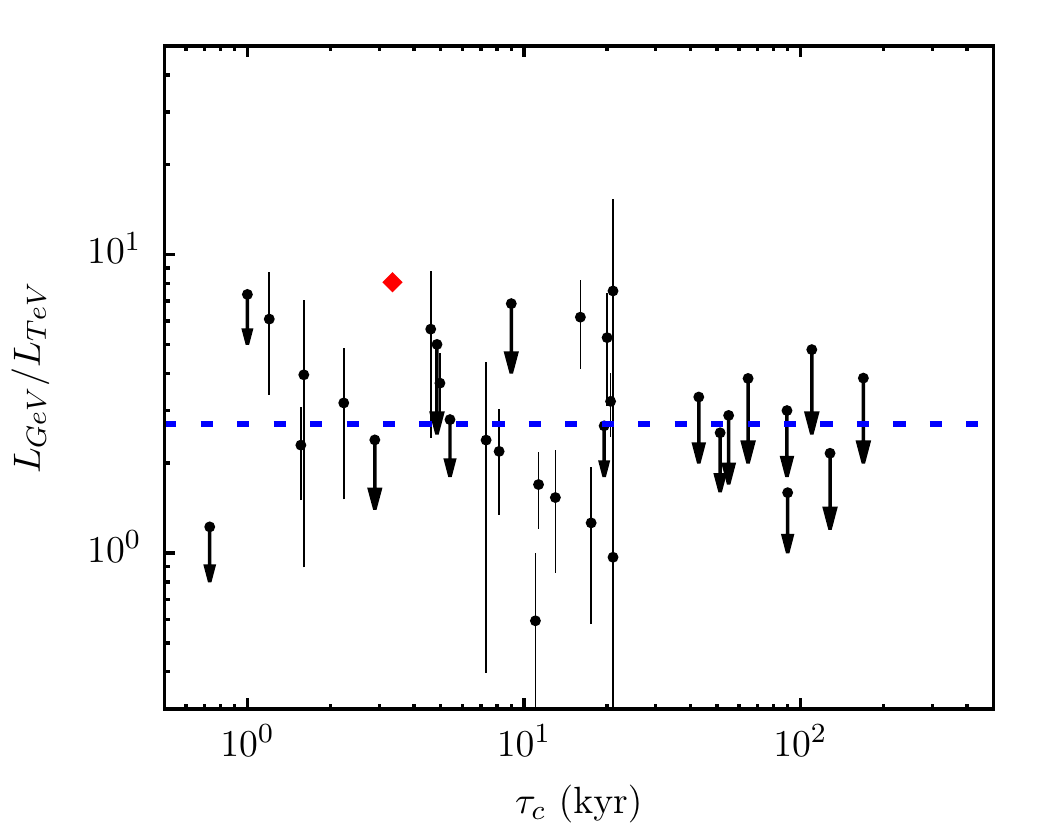}%
\includegraphics[width=0.5\textwidth]{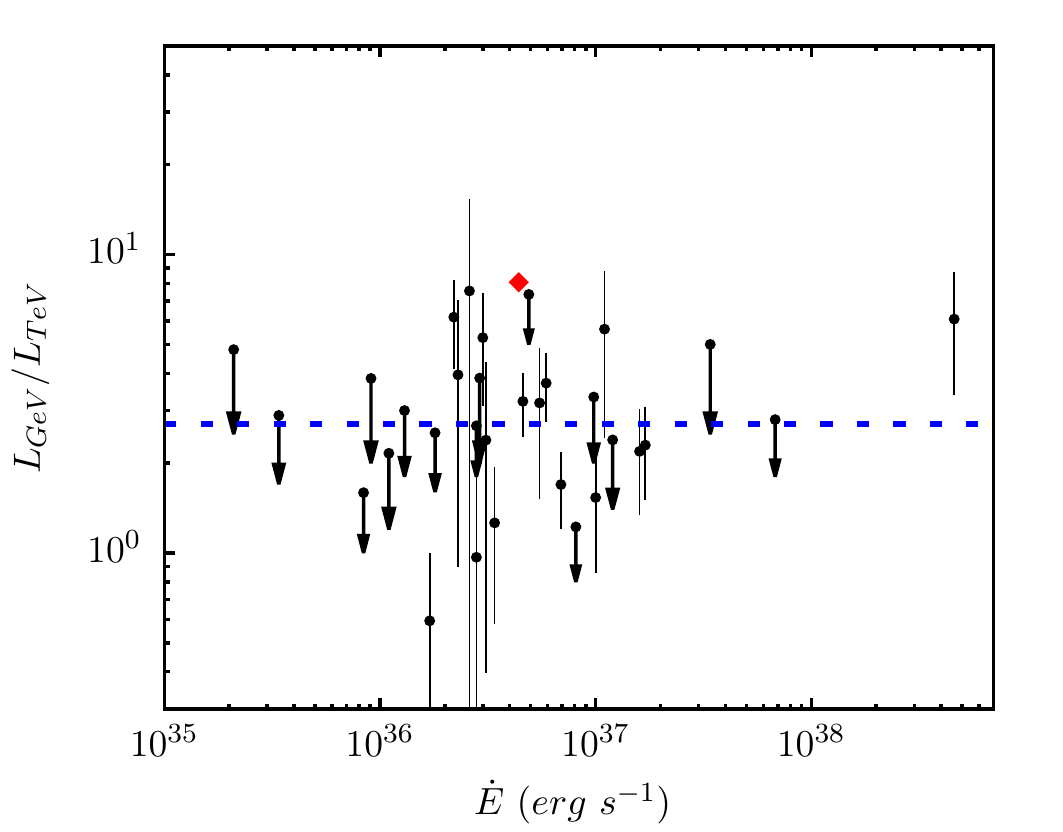}%

\includegraphics[width=0.5\textwidth]{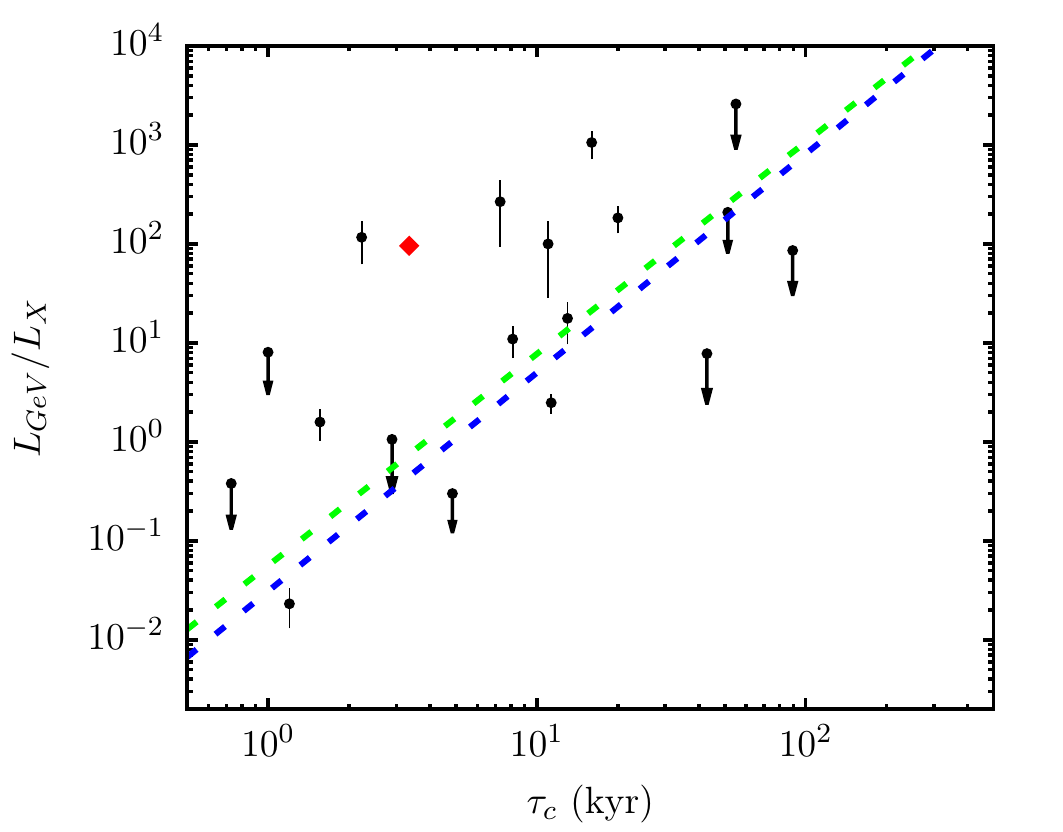}%
\includegraphics[width=0.5\textwidth]{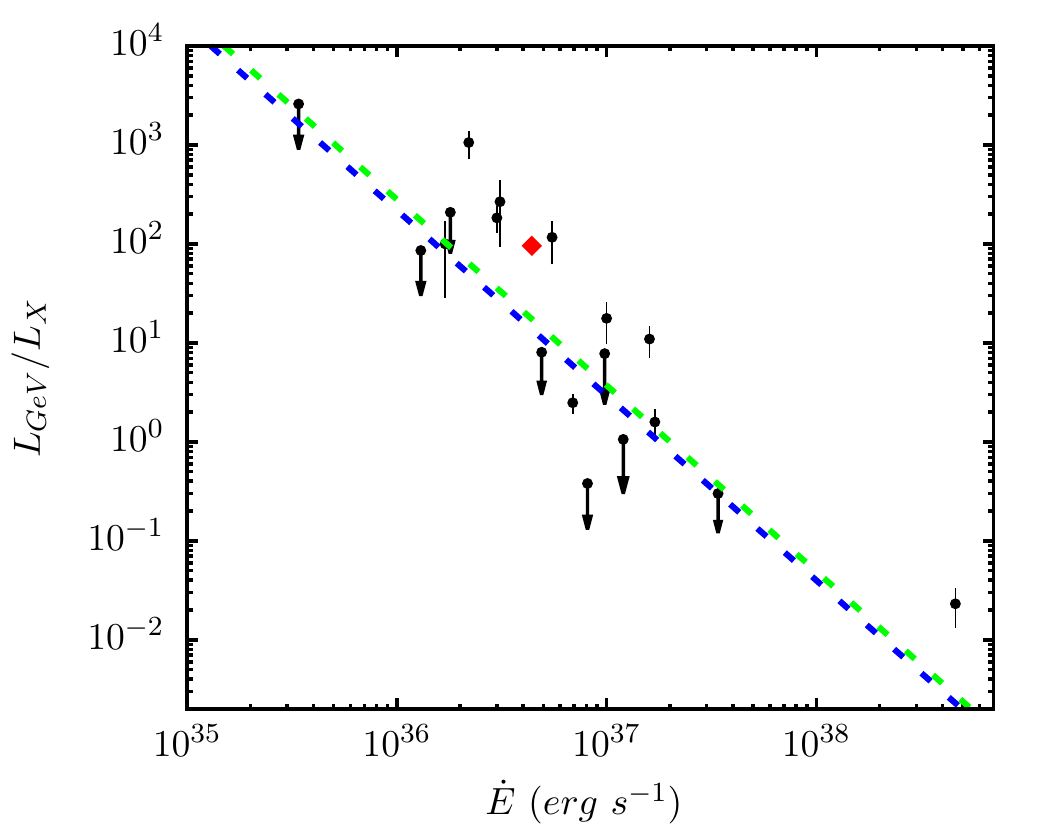}%

\includegraphics[width=0.5\textwidth]{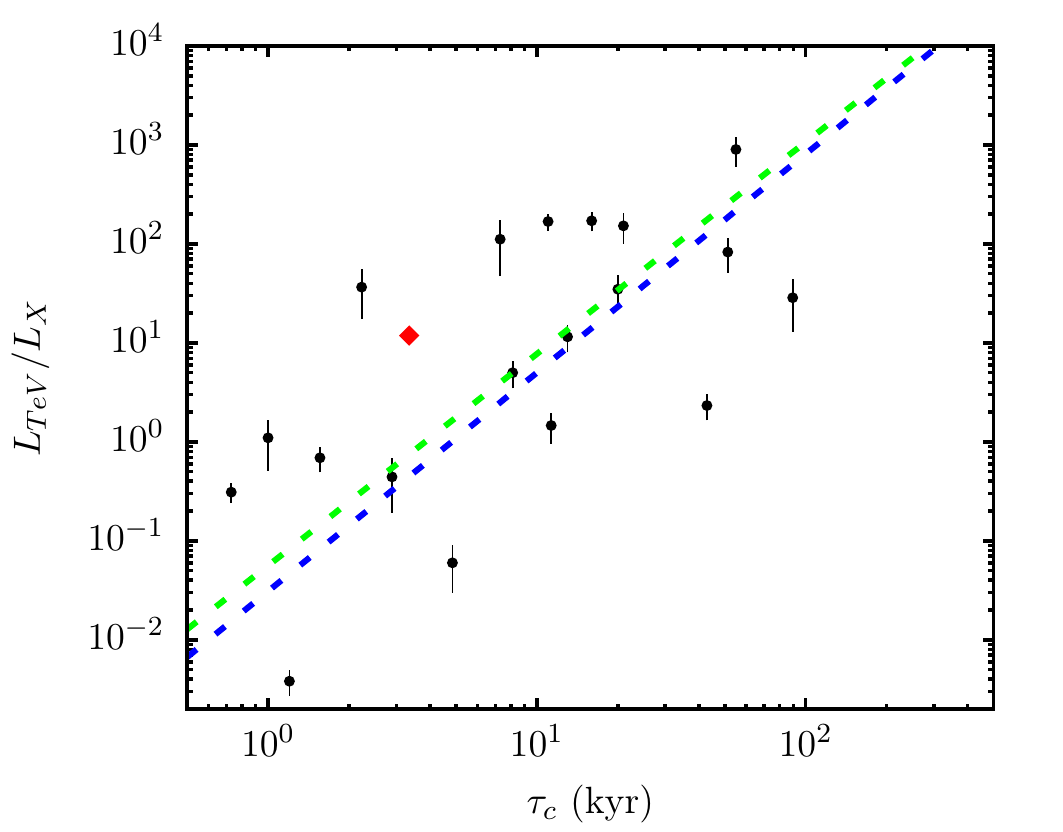}%
\includegraphics[width=0.5\textwidth]{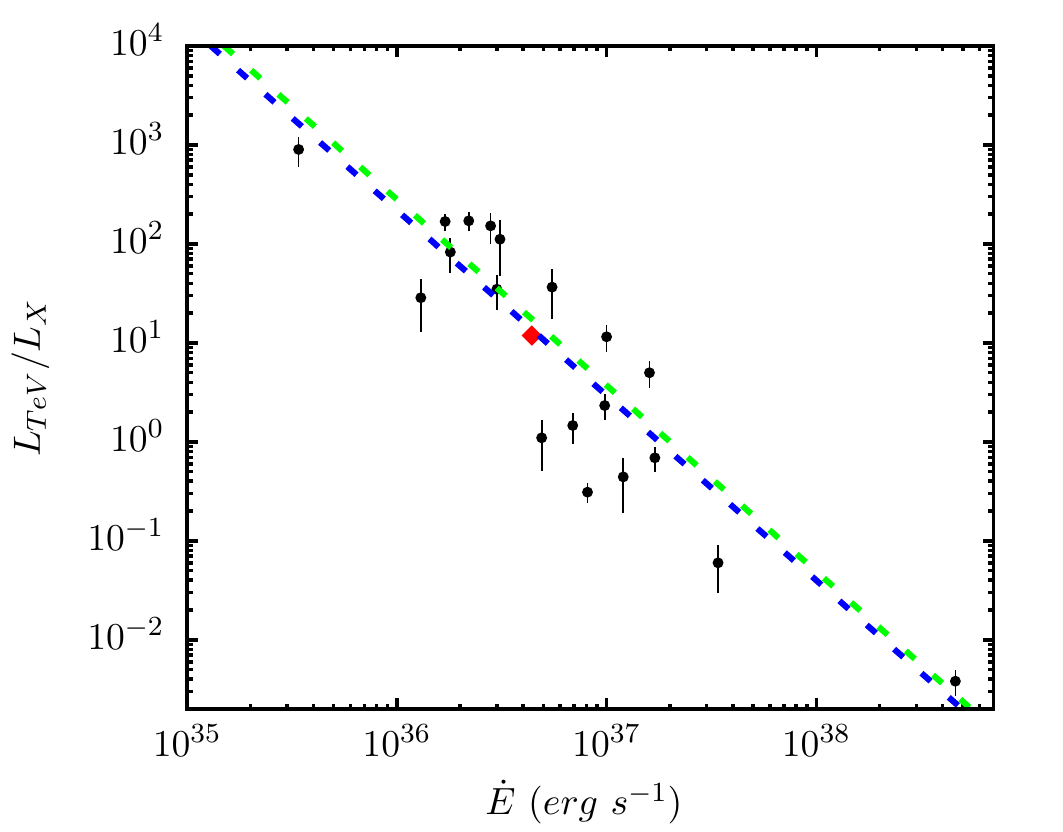}%
\vspace{1em}
\hfill
\caption{From top to bottom: the ratio of GeV to TeV luminosity, $L_{\rm GeV}/L_{\rm TeV}$, 
the ratio of GeV to X-ray luminosity, $L_{\rm GeV}/L_{\rm X-ray}$
and the ratio of TeV to X-ray luminosity, $L_{\rm TeV}/L_{\rm X-ray}$
vs. the characteristic age, $\tau_c$ (left column) 
and the pulsar spin-down luminosity, $\dot{E}$ (right column). 
The X-ray, GeV and TeV luminosities are defined in the energy range of 2-10 keV, 10-316 GeV and 1-30 TeV, respectively.
All data are from \citet{Acero2013}.
The black points represent the sources detected in corresponding energy bands and the arrows are the upper limits.
The red diamond denotes HESS J1640-465, for which, the GeV luminosity of it is adopted to be the result analysed in this work.
The blue dashed lines in the top two panels represent the mean ratio of GeV to TeV luminosity in \citet{Acero2013}.
And in the middle and bottom four panels, the best-fitting statistical relations among the $\gamma$-ray to X-ray luminosity 
ratio ($L_{\rm GeV}/L_{\rm X-ray}$ and $L_{\rm TeV}/L_{\rm X-ray}$) based on the identified PWNe and the whole sample 
in \citet{Mattana2009} are shown as the blue and green dashed lines, respectively.} 
\label{fig:statistic}
\end{figure*}


\citet{Mattana2009} and \citet{Acero2013} studied a large sample of PWNe and investigated the correlations 
between the distance-independent GeV to TeV ($L_{\rm GeV}/L_{\rm TeV}$) and GeV to X-ray ($L_{\rm GeV}/L_{\rm X-ray}$) 
luminosity ratios with the physical properties of the pulsars including the spin-down luminosity, $\dot{E}$ and the characteristic age, $\tau_{c}$.
And the X-ray, GeV and TeV luminosities are integrated in the energy range of 2-10 keV, 10-316 GeV and 1-30 TeV, respectively.
To compare HESS J1640-465 and the other sources associated with PWNe, we plot the different luminosity ratios as a function
of $\tau_{c}$ and $\dot{E}$ following \citet{Mattana2009} and \citet{Acero2013}, which are shown in Fig. \ref{fig:statistic}.
For HESS J1640-465, the characteristic age, $\tau_c$ and the spin-down luminosity $\dot{E}$ are 
$\tau_c \equiv P/2\dot{P} = 3350 ~{\rm yr}$ and $\dot{E} = 4.4 \times 10^{36}$ erg~s$^{-1}$, respectively.
The energy flux from 2 keV to 10 keV of HESS J1640-465 is 
$(5.5\pm0.8) \times 10^{-13}$ erg~cm$^{-2}$~s$^{-1}$ \citep{Gotthelf2014,Lemiere2009}, 
corresponding to the X-ray luminosity of $L_{\rm X-ray} \sim 6.6\times 10^{33}\,(d/10\ {\rm kpc})^2$ erg~s$^{-1}$, assuming a distance of 10kpc.
The GeV $\gamma$-ray luminosity from 10 to 316 GeV is estimated to be 
$L_{\rm GeV} \sim 6.3 \times 10^{35}\,(d/10\ {\rm kpc})^2$ erg~s$^{-1}$
and the TeV $\gamma$-ray luminosity of HESS J1640-465 is 
$L_{>1\mathrm{TeV}} = 7.8\times 10^{34}\,(d/10\ {\rm kpc})^2$ erg~s$^{-1}$ \citep{Abramowski2014a,Abramowski2014aerratum}.
\citet{Acero2013} found that the mean ratio of GeV to TeV luminosity is $2.7_{-1.4}^{+2.7}$.
And for HESS J1640-465, the GeV to the TeV luminosity ratio is located in the 2$\sigma$ region from the mean ratio.

The $\gamma$-ray to X-ray luminosity ratio for the PWNe is found to be proportional to 
the characteristic age of pulsar ($\propto \tau_{c}^{2.2}$), 
but inversely proportional to its 
spin-down luminosity \citep[$\propto \dot{E}^{-1.9}$, equation (3)-(6) in][]{Mattana2009}.
These relations can be interpreted by the time-dependent electron spectrum and the dynamical evolution of 
PWNe \citep{Gelfand2009, Mattana2009, Acero2013}.
As shown in Fig. \ref{fig:statistic}, the statistical parameters of HESS J1640-465
accord with such relations and HESS J1640-465 has the similar properties with the typical PWNe.

\citet{Abdalla2017} systematically studied the population of TeV PWNe found in the H.E.S.S Galactic Plane Survey (HGPS)
and presented the correlations among the properties of TeV PWNe and their respective pulsars, 
such as the TeV extension, the offset between a pulsar and its PWN, the TeV luminosity, 
the TeV surface brightness, the TeV photon index, and so on.
Though HESS J1640-465 was considered as a PWN candidate in \citet{Abdalla2017},
it conforms with all the statistical correlations and the present common time-dependent 
modelling of the typical TeV PWNe, basically \citep{Abdalla2017}.
Overall, HESS J1640-465 does not show obvious differences with the other sources associated with PWNe 
in terms of the correlations among its physical parameters, which makes it more reasonable to be a PWN.

In Figure 5, we also show the spectra of HESS J1427-608, an interesting source with a single power-law spectrum 
in the $\gamma$-ray range. \citet{Guo2017a} show that leptonic models overproduce radio emission of this source.  
However, if one adopts a broken power law model with a spectral index difference greater than 1 as we find for HESS J1640-465, 
the multi-wavelength spectra of HESS J1427-608 can be fit with a model for leptonic emission from an electron population described by a broken power-law distribution in energy.
The fact that it has similar mean magnetic field strength as HESS J1640-465 favors a PWN origin of this source. 

The TeV spectrum of HESS J1641-463 is also very hard, similar to HESS J1427-608, which has been attributed to high-energy 
cosmic ray escaping from G338.3+0.0 by \citet{Tang2015} and \citet{Lau2017}. 
Given the similarity between GeV spectra of source A and HESS J1641-463 and the lack of low energy $\gamma$-ray emission from G338.3-0.0, 
it is possible that source A and HESS J1641-463 are both associated with molecular clouds illuminated 
by a soft cosmic ray flux from G338.3-0.0 \citep{Lau2017}. 
The hard TeV spectrum of  HESS J1641-463 then may be attributed to a PWN in G338.5+0.1. 
We therefore propose a scenario for the complex $\gamma$-ray spectrum of HESS J1641-463, 
which is quite distinct from other previous models \citep{Abramowski2014b, Lemoine-Goumard2014}. 
Future high sensitivity radio and X-ray observations may be able to distinguish these models.

\section{Conclusion}

In this work, we reanalyze the GeV $\gamma$-ray emission in the field of the TeV source, HESS J1640-465, 
using eight years Fermi-LAT data with the latest version of Pass 8.
And two independent sources with different spectral indices, Source A and Source B, 
are detected with significant $\gamma$-ray emission in the energy bands of 1-10 GeV and 10-500 GeV, respectively.
Source A has a soft spectrum with index of $\sim 2.6$ and is treated as a background point source in the field.
Source B is well positionally consistent with HESS J1640-465 and its gamma-ray spectrum is very hard.
The $\gamma$-ray emission of Source B has a spatial extension with a significance level of 5$\sigma$,
which is consist with the TeV $\gamma$-ray extension of HESS J1640-465.
Adopting a 2-D Gaussian profile with $\sigma$ = $4.3'$ as the spatial template of Source B,
the spectral index of Source B in the 10-500 GeV range is found to be $1.42\pm0.19$ for a single power-law spectrum.
And no significant spectral curvature is found for Source B in the GeV band.
The SED of Source B matches well with the TeV spectrum of HESS J1640-465 at a few hundred GeV energies,
which supports Source B as the GeV $\gamma$-ray counterpart of HESS J1640-465.

The best-fitting positions in the GeV and TeV bands of HESS J1640-465 have good coincidence with that of
the PWN powered by PSR J1640-4631.
And the GeV spectrum with index of $1.42\pm0.19$ for HESS J1640-465 is also typical for 
the identified PWNe.
Although several SNRs, such as RX J1713-3946, RX J0852-4622, RCW 86, SN1006, and HESS J1731-347,
show the similarity hard GeV $\gamma$-ray spectra as HESS J1640-465,
all these SNRs emit significant non-thermal X-ray emissions 
that are very different from HESS J1640-465 which no any non-thermal X-ray emission detected from the shell of SNR G338.3-0.0.
Therefore, HESS J1640-465 is more likely to be relevant to the PWN powered by PSR J1640-4631.

We collect the multi-wavelength data of the PWN powered by PSR J1640-4631, and adopting a leptonic model
to constrain the radiation parameters together with the $\gamma$-ray emission of HESS J1640-465.
The synchrotron radiation loss timescale calculated based on the values of the break energy of electrons
and the magnetic field strength in the model is far larger than the characteristic age of PSR J1640-4631,
which indicates that the spectral break should not be from the radiation loss process.
And such break may origin in the initial distribution of injected particles or the acceleration mechanisms switches.
Comparing the calculated total energy of PSR J1640-4631 and the required total energy of electrons above 1 GeV
in the model, a conversion efficiency of about 6\% to 15\% is needed, 
which is reasonable for the typical PWNe.
In addition, HESS J1640-465 also follows the statistical correlations investigated by the sample of known PWNe,
such as the correlations between the luminosity ratios and the physical properties of respective pulsars including 
$\dot{E}$ and $\tau_c$. All these evidences support HESS J1640-465 to be the PWN powered by PSR J1640-4631 
rather than the part shell of the SNR G338.3-0.0.

\section*{Acknowledgments}

This work was supported by the National Key Research and Development Program of China (2016YFA0400200), 
the Key Research Program of Frontier Sciences, CAS, QYZDJ-SSW-SYS024,
the 973 Program of China (2014CB845800), the 100 Talents Program of Chinese Academy of Sciences, 
the National Natural Science Foundation of China (11233001, 11433009, 11761131007, U1738122), 
and the International Partnership Program of Chinese Academy of Sciences (114332KYSB20170008).

\end{document}